\documentclass[aps,pra,twocolumn,superscriptaddress]{revtex4}%
\usepackage{bm}
\usepackage{graphicx}
\usepackage{amsmath}
\usepackage{amsthm}
\usepackage{wasysym}
\usepackage[utf8]{inputenc}
\usepackage{amssymb}
\usepackage{epstopdf}
\usepackage{txfonts}
\usepackage{amsfonts}
\begin{document}
\title{Thermal quantum metrology in memoryless and correlated environments}
\author{Gaetana Spedalieri}
\affiliation{Research Laboratory of Electronics, Massachusetts Institute of Technology,
Cambridge, Massachusetts 02139, USA}
\affiliation{Computer Science \& York Centre for Quantum Technologies, University of York,
York YO10 5GH, United Kingdom}
\author{Cosmo Lupo}
\affiliation{Computer Science \& York Centre for Quantum Technologies, University of York,
York YO10 5GH, United Kingdom}
\author{Samuel L. Braunstein}
\affiliation{Computer Science \& York Centre for Quantum Technologies, University of York,
York YO10 5GH, United Kingdom}
\author{Stefano Pirandola}
\affiliation{Computer Science \& York Centre for Quantum Technologies, University of York,
York YO10 5GH, United Kingdom}

\begin{abstract}
In bosonic quantum metrology, the estimate of a loss parameter is typically
performed by means of pure states, such as coherent, squeezed or entangled
states, while mixed thermal probes are discarded for their inferior
performance. Here we show that thermal sources with suitable correlations can
be engineered in such a way to approach,\ or even surpass, the error scaling
of coherent states in the presence of general Gaussian decoherence. Our
findings pave the way for practical quantum metrology with thermal sources in
optical instruments (e.g., photometers) or at different wavelengths (e.g., far
infrared, microwave or X-ray) where the generation of quantum features, such
as coherence, squeezing or entanglement, may be extremely challenging.

\end{abstract}

\maketitle

\section{Introduction}

Quantum metrology~\cite{SamMETRO,Caves,Paris,Giova,ReviewMETRO,ReviewNEW} is
one of the most active research areas in quantum information
science~\cite{NiCh,Hayashi,Watrous}. The possibility to exploit quantum
resources to boost the estimation of unknown parameters encoded in quantum
states or channels is appealing for a variety of practical tasks, from
gravitational wave detection~\cite{grav1,grav2} to frequency
standards~\cite{freq} and clock synchronization~\cite{clock1,clock2}. In the
specific framework of continuous-variable systems~\cite{WeeRMP,SamRMP},
parameter estimation typically involves the statistical inference of the
phase~\cite{Parisbounds,OlivaresSpe,Paris07,GenoniParis,GenoniParis2,Dowling2,Dowling3,GenoniKim,Raf0,Escher,Yue,Raf1,Raf2,Bagan}
or
loss~\cite{MonrasParis,Treps,Venzel,AdessoIlluminati,Gaiba,MonrasIlluminati}
accumulated by a bosonic mode propagating through a Gaussian channel. For this
task, Gaussian and non-Gaussian resources have been extensively
studied~\cite{ReviewNEW}.

While the minimization of the estimation error over all quantum strategies is
crucial to show the ultimate precision achievable by quantum mechanics, it is
also important to study practical applications to realistic scenarios, where
the access to quantum resources may be limited and the presence of decoherence
may even destroy the quantum advantage shown for the noiseless models. This is
an important gap to fill for bosonic systems, where previous studies on loss
estimation were devoted to finding the optimal error scaling reachable by
squeezing, entanglement or other highly non-classical features in
decoherence-free scenarios~\cite{MonrasParis,Treps,Venzel,Gaiba}.

Here we extend the state-of-the-art on bosonic loss estimation in two ways.
First of all, we consider the practical use of correlated-thermal sources
which can be easily engineered by using a beam splitter. These sources are
generally designed to be asymmetric so that only a few mean photons are
irradiated through the unknown lossy channel, while the majority of them are
deviated onto an ancillary channel. Thanks to this asymmetric splitting, the
lossy channel is `non-invasively' probed with low energy, while enough
correlations are created with the ancillary photons to improve the final detection.

This practical scheme is relevant in various realistic scenarios. For
instance, this is the simplest strategy to improve the optical setups of
photometers and spectrophotometers currently employed in experimental biology.
These instruments use thermal lamps at optical or UV wavelengths to measure
the concentration of bacteria, cells, or nucleic acids (DNA/RNA) in fragile
biological samples via an estimation of the transmissivity~\cite{Ingle}. Our
interferometric design would introduce correlations and greatly improve their performance.

Other important scenarios are the far infrared and microwave regimes where
quantum features are hard to generate. By contrast, correlated-thermal sources
can be easily generated in these cases, and could be adopted (in the long run)
to advance applications such as protein Terahertz spectroscopy or magnetic
resonance imaging. Similar implications could also be envisaged at very
high-frequencies, where quasi-monochromatic X-ray beams can now be generated
by small-scale all-laser-driven Compton sources with good\ spatial-temporal
coherence~\cite{Powers}. These thermal beams could be manipulated by X-ray
beam splitters based on Laue-Bragg diffraction~\cite{Oberta} or other X-ray
interferometry~\cite{Nugent}.

Besides the focus on cheap correlated-thermal sources, the second novelty of
our work is to provide the first study of loss estimation assuming a general
model of Gaussian decoherence, which includes additional loss, thermal effects
and even the possibility of environmental correlations. Thanks to this general
model, we can potentially account for many effects, including detector
inefficiencies, thermal background (which is non-trivial at the microwave
regime) and also the presence of non-Markovian dynamics in the environment.

In such a general scenario, we fix the benchmark to be the performance of
coherent states: The generation of minimum uncertainty states can be regarded
as the minimal requirement for a single-mode source to be considered
`quantum'. While the direct use of single-mode thermal sources is clearly
sub-optimal, we show that the coherent-state benchmark can easily be achieved
by two-mode thermal sources which are asymmetric and correlated. Surprisingly,
these sources are even able to largely outperform the coherent-state benchmark
when (separable) correlations are present in the environment.

\section{Quantum Metrology with Correlated-Thermal Sources}

\begin{figure}[ptbh]
\vspace{-1.3cm}
\par
\begin{center}
\includegraphics[width=0.49\textwidth] {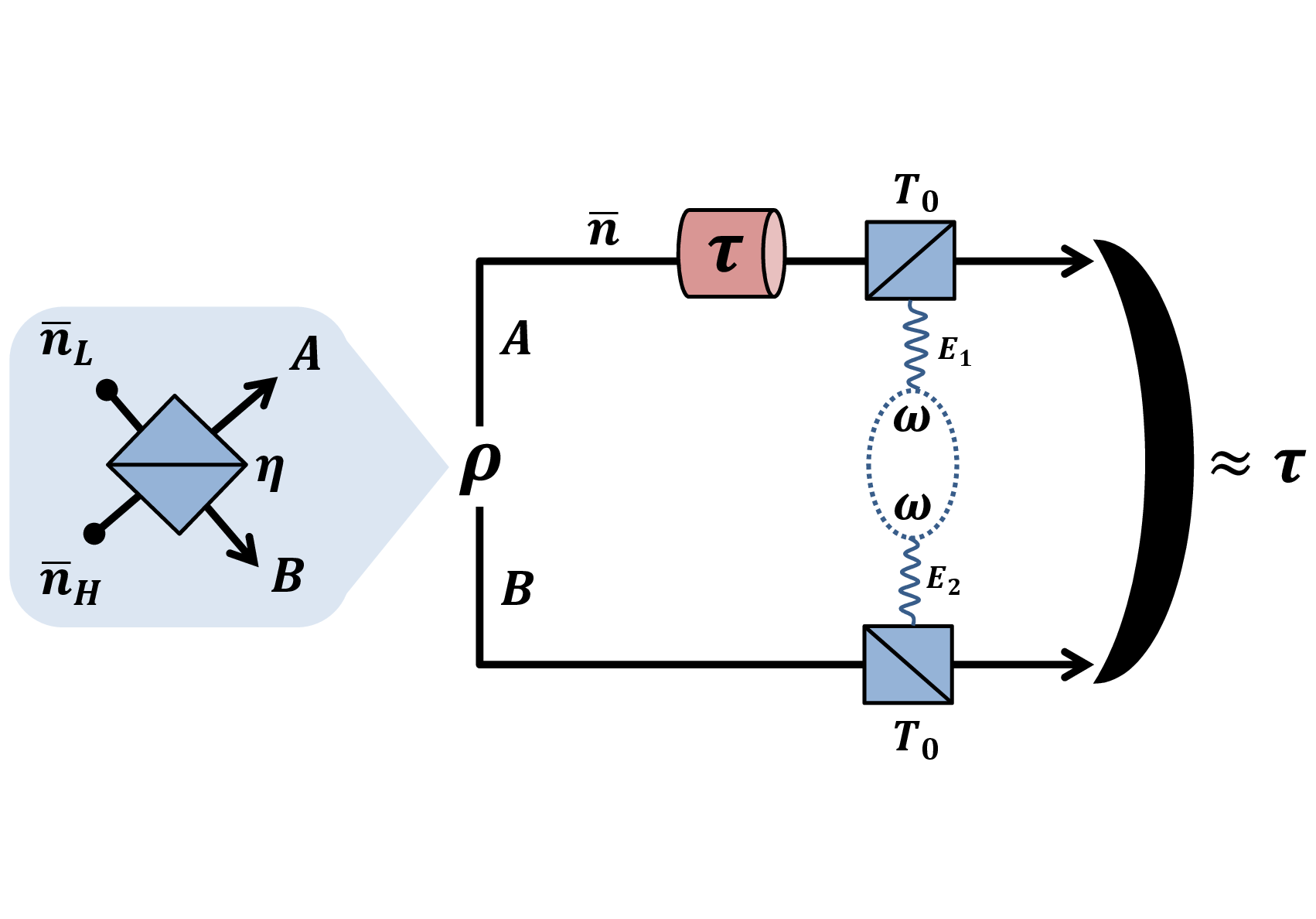}
\end{center}
\par
\vspace{-1.6cm}\caption{Bosonic loss estimation with correlated-thermal
sources under general Gaussian decoherence. On the left we show the
preparation of a two-mode thermal source $\rho_{AB}$ which is correlated and
generally asymmetric. This is prepared using a beam splitter with
transmissivity $\eta\leq1/2$, which mixes a bright thermal state (with photon
number $\bar{n}_{H}$) with a faint thermal state (with photon number $\bar
{n}_{L}<$ $\bar{n}_{H}$). The three parameters of the source ($\eta$, $\bar
{n}_{H}$ and $\bar{n}_{L}$) are chosen in such a way that the mean number of
photons $\bar{n}$ on mode $A$ is fixed to some low value, while the energy of
mode $B$ is not constrained. On the right, the input thermal source and an
optimal output measurement are employed\ for estimating the unknown
transmissivity $\tau$ of the red box (lossy channel). This is done in the
presence of Gaussian decoherence, modelled by two beam splitters with
transmissivity $T_{0}$ and injecting thermal noise with variance $\omega$. The
environmental thermal modes $E_{1}$ and $E_{2}$ may be uncorrelated or
correlated.}%
\label{basicPIC}%
\end{figure}

Let us start with a detailed description of the correlated-thermal source (see
also Fig.~\ref{basicPIC}). We consider two single-mode thermal states,
$\rho_{H}$ and $\rho_{L}$, with mean numbers of photons equal to $\bar{n}_{H}$
and $\bar{n}_{L}$, respectively. These are chosen to satisfy $\bar{n}_{H}%
>\bar{n}_{L}$ and we may specifically consider $\bar{n}_{L}=0$. The two
thermal states are combined with a generally unbalanced beam splitter, with
transmissivity $\eta\leq1/2$. The three parameters of the source ($\eta$,
$\bar{n}_{H}$ and $\bar{n}_{L}$) are chosen in such a way that the mean number
of photons transmitted on mode $A$, equal to $\bar{n}=\eta\bar{n}_{H}%
+(1-\eta)\bar{n}_{L}$, is fixed to some low value (e.g., $\bar{n}=10$), while
no energetic constraint is imposed for mode $B$.

As mentioned above, the most interesting situation is when the source is
highly asymmetric. This means that we take $\bar{n}_{H}\gg\bar{n}_{L}\simeq0$
and $\eta\ll1$, in such a way that $\bar{n}\simeq\eta\bar{n}_{H}$ is kept
small, while mode $B$ is very energetic with $\simeq\bar{n}_{H}$ photons
transmitted. Locally, the reduced state $\rho_{A}$ ($\rho_{B}$) is a faint
(bright) thermal state, but globally the state $\rho_{AB}$ is highly
correlated. One can check that the quadrature operators associated with the
two modes ($\hat{q}_{A}$,$\hat{p}_{A}$, $\hat{q}_{B}$ and $\hat{p}_{B}$) have
covariances $\left\langle \hat{q}_{A}\hat{q}_{B}\right\rangle =\left\langle
\hat{p}_{A}\hat{p}_{B}\right\rangle \simeq-\bar{n}\eta^{-1/2}$, whose absolute
value is $\gg1$. 

The generated thermal source $\rho_{AB}(\eta,\bar{n}_{H},\bar{n}_{L})$ is then
used to probe a lossy channel $\mathcal{E}_{\tau}$ with unknown transmissivity
$\tau\in\lbrack0,1]$ on mode $A$. In a realistic scenario, this is affected by
decoherence, here modelled by a generally-joint Gaussian channel $\mathcal{D}$
affecting both modes $A$ and $B$. This can be represented by two beam
splitters with transmissivity $T_{0}$ mixing $A$ and $B$ with ancillary modes,
$E_{1}$ and $E_{2}$, coming from the environment. These ancillas inject
thermal noise $\omega=\bar{n}_{\text{env}}+1/2$, where $\bar{n}_{\text{env}}$
is the mean number of photons of the bath. Furthermore, the two environmental
ancillas may also be correlated, which means that their quadrature operators,
i.e., $\hat{q}_{E_{1}}$,$\hat{p}_{E_{1}}$, $\hat{q}_{E_{2}}$ and $\hat
{p}_{E_{2}}$, have non-zero covariance, i.e., $\left\langle \hat{q}_{E_{1}%
}\hat{q}_{E_{2}}\right\rangle =g$ and $\left\langle \hat{p}_{E_{1}}\hat
{p}_{E_{2}}\right\rangle =g^{\prime}$, satisfying suitable
constraints~\cite{CMbona,CMbona2} (see Appendix~\ref{app3}). Thus, the output
Gaussian state is given by $\rho_{AB}^{\text{out}}(\tau)=\mathcal{D}%
\circ(\mathcal{E}_{\tau}\otimes\mathcal{I})(\rho_{AB})$.

At the output a joint quantum measurement $\mathcal{M}$ is performed on modes
$A$ and $B$ whose outcome provides an estimate of $\tau$. In the basic
formulation of quantum metrology, this process is assumed to be performed many
times, so that a large number $N\gg1$ of input states $\rho_{AB}^{\otimes N}$
are prepared and their outputs $\rho_{AB}^{\text{out}}(\tau)^{\otimes N}$ are
subject to a collective quantum measurement $\mathcal{M}^{\otimes N}$, whose
output is classically processed into an unbiased estimator $\tilde{\tau}_{N}$
of $\tau$. For large $N$, the resulting error-variance $\sigma^{2}%
(\tau,N):=\langle(\tilde{\tau}_{N}-\tau)^{2}\rangle$ satisfies the quantum
Cramer-Rao (QCR) bound $\sigma^{2}(\tau,N)\geq\left[  NH(\tau)\right]  ^{-1}$,
where $H(\tau)$ is the quantum Fisher information (QFI)~\cite{SamMETRO}. The
QFI can be expressed as $H(\tau)=8(1-F)/d\tau^{2}$, where $F$ is the quantum
fidelity between the two Gaussian states $\rho_{AB}^{\text{out}}(\tau)$ and
$\rho_{AB}^{\text{out}}(\tau+d\tau)$, which can be computed using the general
formula of Ref.~\cite{Benki}. It is important to note that the QCR bound can
always be achieved, asymptotically, by an optimal measurement $\mathcal{M}%
^{\otimes N}~$\cite{SamMETRO}.

In the following we show the performances achievable by our correlated-thermal
sources under various assumptions for the Gaussian decoherence model, starting
from the simple case of a pure-loss environment, to including thermal noise
and, finally, noise-correlations. These performances are compared with the use
of a single-mode thermal source and, most importantly, with a coherent-state
benchmark. The latter can easily be evaluated. Considering the scenario at the
right of Fig.~\ref{basicPIC}, but neglecting mode $B$ and considering an input
coherent state $\left\vert \alpha\right\rangle $ with $\left\vert
\alpha\right\vert ^{2}=\bar{n}$ on mode $A$, we derive the benchmark (see
Appendix~\ref{app2} for details)%
\begin{equation}
H_{\text{coh}}(\tau)=\frac{\gamma_{\text{dec}}\bar{n}}{\tau},~~~~\gamma
_{\text{dec}}:=\frac{T_{0}}{T_{0}+2(1-T_{0})\omega}. \label{bench}%
\end{equation}
In this formula, we can see how the error-scaling $\varpropto\bar{n}/\tau$ is
moderated by the factor $\gamma_{\text{dec}}$ taking into account of the
Gaussian decoherence.

\section{Pure-loss decoherence}

Let us start with the simplest decoherence model, which only considers
additional damping on top of the unknown lossy channel under estimation. In
other words, we consider the two beam splitters with $T_{0}<1$ in a zero
temperature bath ($\omega=1/2$) and without noise correlations ($g=g^{\prime
}=0$). This is the most typical situation at the optical regime, where thermal
background is negligible. Such a pure-loss decoherence may be found in many
scenarios. For instance, it may be the effect of detector inefficiencies, beam
spreading, or the use of fiber components. In other cases, it may due to the
typical configuration of an optical instrument. For example, in a photometer,
the measure of a concentration within a sample (via its optical transmission)
is typically performed with respect to a blank sample whose intrinsic
transmissivity is known and fixed.

\begin{figure}[ptbh]
\vspace{-0.2cm}
\par
\begin{center}
\includegraphics[width=0.46\textwidth] {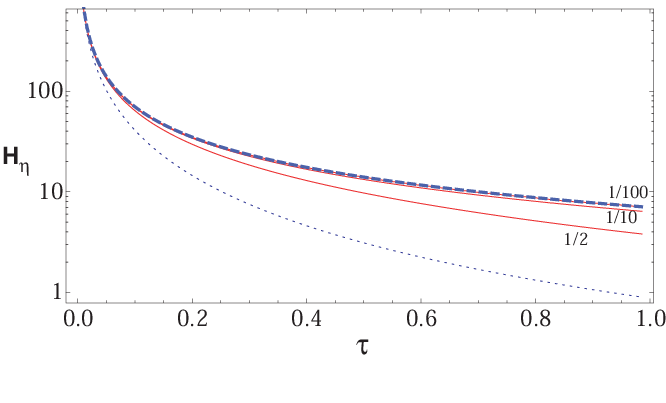}
\end{center}
\par
\vspace{-0.99cm}\caption{Quantum Fisher information $H_{\eta}$ versus
transmissivity $\tau$ for probes irradiating $\bar{n}=10$ signal photons. We
plot the performances of the correlated-thermal source for $\eta=1/2$, $1/10$
and $1/100$ (solid red lines); larger $H_{\eta}$ indicates better precision.
These are compared with\ the single-mode thermal state (dotted blue line) and
the coherent-state benchmark (dashed blue line, which coincides with the solid
red line for $\eta=1/100$). Here we consider $T_{0}=0.7$, $\omega=1/2$ (zero
temperature bath), and $g=g^{\prime}=0$ (corresponding to no correlations in
the environment).}%
\label{PIC1}%
\end{figure}

Let us estimate the transmissivity parameter $\tau$\ by constraining the mean
number of photons in the signal mode $A$, e.g., $\bar{n}=10$, and assuming
additional (known) loss in modes $A$ and $B$, e.g., quantified by $T_{0}=0.7$.
We then construct correlated-thermal sources combining a strongly attenuated
thermal state $\bar{n}_{L}=10^{-4}\simeq0$ (approximately the vacuum
state)\ and a thermal state with $\bar{n}_{H}=\bar{n}\eta^{-1}$, where the
parameter $\eta$\ of the beam splitter is variable and completely describes
the source. The corresponding QFI $H_{\eta}(\tau)$ is plotted in
Fig.~\ref{PIC1}, where the performances of these sources are compared with
that of the single-mode thermal state (achievable by setting $\eta=1$) and
that of the coherent state probes, according to Eq.~(\ref{bench}) with
$\gamma_{\text{dec}}=T_{0}$. 

As we can see from Fig.~\ref{PIC1}, the correlated-thermal source is optimal
in the most asymmetric configurations, where the beam splitter is highly
unbalanced (e.g., $\eta=1/100$) so that strong correlations are generated
between the signal mode $A$ and the ancillary mode $B$, while keeping the
signal energy low at $\bar{n}=10$ photons. The coherent-state benchmark is
easily approached already with reasonable asymmetries (e.g., $\eta=1/10$). It
is remarkable that the performance achievable by coherent photons on mode $A$
can also be achieved by employing an equivalent number of thermal photons (as
long as they are suitably correlated with the ancillary mode $B$).

Note that highly-asymmetric beam splitters are typical in X-ray
interferometry. A hard X-ray beam at $25~$KeV (suitable for medical
applications, such as mammography) can be split by Silicon crystals via
Laue--Bragg diffraction. For crystals of sufficient depth $\simeq200~\mu$m,
the diffraction efficiency (reflectivity of the beam splitter) can reach
values of $1/100$~\cite{Oberta}.

\section{Thermal-loss decoherence}

\begin{figure}[ptbh]
\vspace{-.18cm}
\par
\begin{center}
\includegraphics[width=0.46\textwidth] {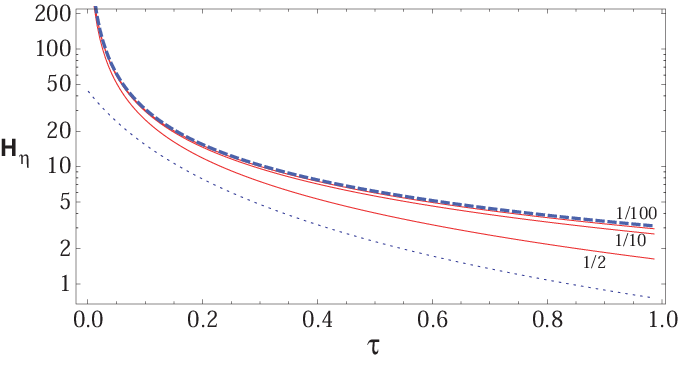}
\end{center}
\par
\vspace{-0.55cm}\caption{Quantum Fisher information $H_{\eta}$ versus
transmissivity $\tau$ for probes irradiating $\bar{n}=20$ signal photons. We
plot the performances of the correlated-thermal source with $\bar{n}_{L}%
\simeq0.12$ and having $\eta=1/2$, $1/10$ and $1/100$ (solid red lines);
larger $H_{\eta}$ indicates better precision. These are compared with\ the
single-mode thermal state (dotted blue line) and the coherent-state benchmark
(dashed blue line). Here we consider $T_{0}=0.4$, $\omega\simeq1.33+1/2$ (room
temperature at $3.5$ THz), and $g=g^{\prime}=0$ (corresponding to no
correlations in the environment).}%
\label{PIC2}%
\end{figure}

We now include the presence of thermal noise in the decoherence process.
Besides various technical imperfections (e.g., stray photons emitted by the
source), this noise may come from a natural thermal background which is
non-negligible at far infrared and microwave wavelengths. As an example,
consider the frequency of $3.5$ THz. At room temperature ($300$ K) there will
be $\bar{n}_{\text{env}}\simeq1.33$\ mean thermal photons entering the
interferometric setup (via the input ports $E_{1}$ and $E_{2}$ of the two beam
splitters of Fig.~\ref{basicPIC}). Assuming a liquid-nitrogen temperature
($77$ K) for the preparation beam splitter, we have $\bar{n}_{L}\simeq0.12$.
We then consider high loss ($T_{0}=0.4$) and signals with $\bar{n}=20$
photons. As we can see from Fig.~\ref{PIC2}, correlated-thermal sources with
enough asymmetry are again able to approach the coherent-state
benchmark.

\section{Correlated-noise decoherence}

We finally consider noise correlations in the Gaussian environment. There may
be situations, e.g., on a small scale, where two bosonic modes experience
exactly the same fluctuations. In these `non-Markovian' environments, we find
that our correlated-thermal sources can even beat the coherent-state
benchmark. More specifically, we consider noise-correlations of the type
$g=g^{\prime}=1/2-\omega\leq0$, which are maximal but still separable (i.e.,
the state of the environmental ancillas $E_{1}$ and $E_{2}$ is not entangled).
These specific environmental correlations constructively combine with those of
the input two-mode thermal source in a way as to reduce the net effect of decoherence.

\begin{figure}[ptbh]
\vspace{+0.15cm}
\par
\begin{center}
\includegraphics[width=0.46\textwidth] {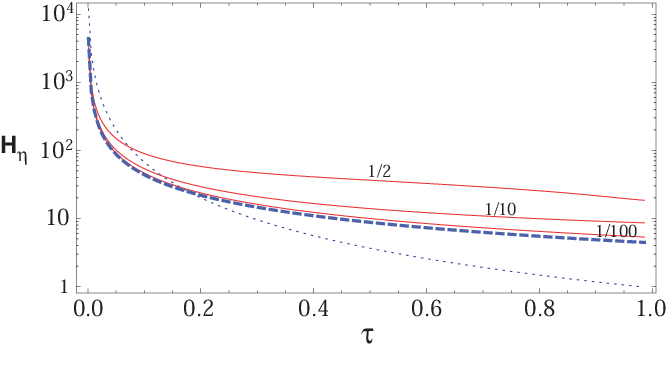}
\end{center}
\par
\vspace{-0.5cm}\caption{Quantum Fisher information $H_{\eta}$ versus
transmissivity $\tau$ for probes irradiating $\bar{n}=50$ signal photons. We
plot the performances of the correlated-thermal source with $\bar{n}_{L}%
\simeq8.3\times10^{-3}$ and having $\eta=1/2$, $1/10$ and $1/100$ (solid red
lines); larger $H_{\eta}$ indicates better precision. These are compared
with\ the single-mode thermal state (dotted blue line) and the coherent-state
benchmark (dashed blue line). Here $T_{0}=0.8$, $\omega\simeq20.34+1/2$ and
$g=g^{\prime}=1/2-\omega$.}%
\label{PIC3}%
\end{figure}

Consider $T_{0}=0.8$ and $\bar{n}_{\text{env}}\simeq20.34$ (e.g.,
corresponding to $300$ GHz at room temperature). We shall assume we have
correlated-thermal sources with $\bar{n}_{L}\simeq8.3\times10^{-3}$ (e.g., via
a cryogenic preparation), variable $\eta$, and irradiating $\bar{n}=50$
photons on mode $A$. As we can see from Fig.~\ref{PIC3}, all choices of
sources beat the coherent-state benchmark. In particular, the best solution is
the fully-symmetric thermal source with $\eta=1/2$, which is the most
effective at counterbalancing the specific symmetric noise of the environment.
We can easily extend this analysis to considering an environment with
asymmetric thermal noise, in which case the best performance is achieved by
asymmetric thermal sources. This is shown in Appendix~\ref{app4}, where we
also check that the coherent benchmark is not beaten if the environmental
correlations are of the \textquotedblleft positive type\textquotedblright%
\ $g=g^{\prime}>0$, therefore not sustaining those \textquotedblleft
negative\textquotedblright\ $\left\langle \hat{q}_{A}\hat{q}_{B}\right\rangle
=\left\langle \hat{p}_{A}\hat{p}_{B}\right\rangle <0$\ of the input correlated
state.

\section{Conclusions}

We have shown that thermal sources can be engineered in such a way that their
correlations may non-trivially improve the performance of loss estimation in
practical setups of quantum metrology considering various scenarios of
Gaussian decoherence. Correlated-thermal sources with strong energetic
asymmetry are able to approach the performance of coherent state probes in
Markovian (memoryless) models of Gaussian decoherence, where the bosonic modes
are affected by independent and identical noise fluctuations. In the presence
of correlated noise in the environment, as typical of non-Markovian dynamics,
we have shown that the correlated-thermal sources can even beat the coherent
state benchmark, a feature which may be achieved by correctly combining the
types of correlations created in the source with those present in the environment.

According to our investigations, the behavior represented in the previous
numerical examples is generic as long as the mean number of photons on mode
$A$ is reasonably low, and the preparation of the correlated-thermal source
involves a faint thermal state with sufficiently low thermal number $\bar
{n}_{L}$ (ideally, this should be the vacuum state). One may argue that the
correlations employed in our sources still have a quantum component, e.g., as
quantified by quantum discord~\cite{DiscordRMP} (which is exactly computable
for these types of Gaussian states~\cite{OptimalityDiscord}). In this respect,
we notice that discord may be considered as the cheapest non-classical feature
to be generated in a bipartite source. Indeed, in our case, it just
corresponds to the ability of combining thermal states at a beam splitter,
which just requires sufficient spatial-temporal coherence in the bosonic
modes. Clearly, this is much less demanding than the ability to generate
minimum uncertainty states or even squeezing.

Further work includes the analysis of loss estimation with a finite number of
signals, and the design of explicit detection strategies able to approach the
theoretical performance of the quantum Cramer-Rao bound. For the case of a
pure-loss environment ($\omega=1/2$), we provide this study in
Appendix~\ref{APP_receivers}, where we show that photon-counting applied to a
correlated-thermal source achieve the same energy scaling in $\bar{n}$ of the
coherent-state benchmark (but with a different pre-factor).

Another potential investigation is considering adaptive strategies for loss
estimation, whose optimal perfomance is unknown. A possible methodology to
exploit is that of channel simulation recently developed in quantum
metrology~\cite{ReviewMETRO,PirCo,PBTstretch} after successful applications in
quantum and private communications~\cite{PLOB,TQC}. It would also be very
interesting to analyze the explicit performance of correlated-thermal sources
for practical tasks of quantum hypothesis
testing~\cite{QHT,QHT1,QHT2,UNA1,UNA2,UNA3,UNA4,UNA5,Invernizzi}, such as the
quantum reading of optical
memories~\cite{Qread,Qread2,Qread3,Qread4,Qread4b,Qread5,Qread6,Qread7,Qread7b,Qread7c,Qread8,Qread9,Qread9b,Qread10}
and the quantum illumination of
targets~\cite{Qill0,Qill1,Qill1b,Qill2,Qill3,Qillexp1,Qillexp2,Qillexp3}.

\section*{Acknowledgments}

G.S. acknowledges the support of the European Commission under the Marie
Sklodowska-Curie Fellowship Progamme (EC\ grant agreement No 745727). S. P.
was supported by the Leverhulme Trust (`qBIO' fellowship) and the EPSRC
(`qDATA', EP/L011298/1).

\appendix

\section{Relations with previous literature\label{app1}}

Previous literature on quantum metrology with bosonic systems has been devoted
to the estimation of various parameters of a Gaussian channel, including
displacement, phase-shift, loss and thermal parameters. Optimal estimation of
displacements was studied in Ref.~\cite{GenoniJoint}. Phase estimation has
undergone an extensive analysis: Bounds on the precision of phase-estimation
using Gaussian resources were studied in Ref.~\cite{Parisbounds}; optimized
interferometry for phase estimation was given in Ref.~\cite{OlivaresSpe}; and
the use of squeezing in high-sensitive interferometry was analyzed in
Refs.~\cite{Paris07,Dowling2,Dowling3}. Phase estimation was also extended to
the presence of decoherence, mainly phase diffusion and photon loss. For
instance, it was extended to phase diffusion in
Refs.~\cite{GenoniParis,GenoniParis2}, to unitary and random linear
disturbance in Ref.~\cite{GenoniKim}, and to lossy optical interferometry in
Refs.~\cite{Raf0,Escher,Yue}, with associated general studies of quantum
metrology with uncorrelated noise~\cite{Raf1,Raf2}. Phase estimation with
displaced thermal states and squeezed thermal state was studied in
Ref.~\cite{Bagan}\ also considering the presence of loss.

There is relatively less literature regarding the estimation of the loss
parameter. Optimal estimation of loss in Gaussian channels was studied in
Ref.~\cite{MonrasParis} by using single-mode pure Gaussian states (see also
Ref.~\cite{Treps}). This analysis was also carried out for entangled Gaussian
states in Ref.~\cite{Venzel} and non-Gaussian sources in
Ref.~\cite{AdessoIlluminati}. All these studies were performed in the absence
of decoherence. Use of squeezing for estimating the interaction parameter in
bilinear bosonic Hamiltonians (including beam-splitter interactions) was also
discussed in Ref.~\cite{Gaiba}. Later, Ref.~\cite{MonrasIlluminati} considered
the joint estimation of damping and temperature of a Gaussian channel by means
of Gaussian states, specifically showing the superior performances achieved by
the use of entanglement over coherent states. Note that this study considered
the presence of thermal noise directly in the single-mode Gaussian channel
under estimation, not the presence of a lossy and thermal environment
(affecting signal and ancillary modes) on top of the channel to be estimated.
Finally, the detection of loss by using squeezed thermal states (in absence of
decoherence) was analyzed in the framework of quantum hypothesis
testing~\cite{Invernizzi}.

Our work departs from all this previous literature in several aspects. First
of all, it is clearly not related with the extensive literature on phase
estimation, since we are considering the loss parameter. Then, with respect to
previous works on bosonic damping estimation, we are:

\bigskip

\noindent\textbf{(i)}~Engineering new correlated-type of thermal states, void
of squeezing, and never investigated before for quantum metrology tasks. These
cheap sources are important for making quantum metrology practical, especially
when considering longer wavelengths, where quantum features are challenging to generate.

\bigskip

\noindent\textbf{(ii)}~Considering a general model of Gaussian decoherence
affecting both signal and ancillary modes, which may introduce loss, thermal
noise, and even two-mode correlations. It is important to note that this type
of decoherence is added on top of the lossy channel to be estimated, and may
describe various realistic scenarios, such as the detector inefficiencies,
thermal background (e.g., at the microwaves), and potential non-Markovian effects.

\section{Coherent-state benchmark\label{app2}}

First of all, a brief remark on the notation. We consider quadrature operators
$\hat{q}$ and $\hat{p}$ with canonical commutation relations $[\hat{q},\hat
{p}]=i$, so that the annihilation operator corresponds to $\hat{a}=(\hat
{q}+i\hat{p})/\sqrt{2}$ and the vacuum shot-noise is equal to $1/2$.
Correspondingly, the covariance matrix (CM) of a single-mode thermal state is
equal to $\mu\mathbf{I}$, where $\mu=\bar{n}+1/2$, with $\bar{n}$ being the
mean number of thermal photons. For the general formalism of
continuous-variable systems and Gaussian states, the reader may consult the
reviews of Refs.~\cite{WeeRMP,SamRMP}.

Let us prepare mode $A$ in a coherent state $\left\vert \alpha\right\rangle
$\ with $\bar{n}=\left\vert \alpha\right\vert ^{2}$ mean photons. This state
has mean value $\mathbf{\bar{x}}=(\bar{q},\bar{p})^{T}$ where $\alpha=(\bar
{q}+i\bar{p})/\sqrt{2}$\ and covariance matrix (CM) equal to $\mathbf{I}/2$.
This is subject to the action of the lossy channel $\mathcal{E}_{\tau}$
followed by that of the thermal-loss decoherence channel $\mathcal{D}_{A}$
with transmissivity $T_{0}$ and thermal noise $\omega$. At the output, the two
statistical moments of $\rho_{\alpha}(\tau,T_{0},\omega)=(\mathcal{D}_{A}%
\circ\mathcal{E}_{\tau})(\left\vert \alpha\right\rangle \left\langle
\alpha\right\vert )$ are given by $\mathbf{\bar{x}}^{\prime}=\sqrt{T_{0}\tau
}\mathbf{\bar{x}}$ and $\mathbf{V}^{\prime}=a^{\prime}\mathbf{I}$, where%
\begin{equation}
a^{\prime}:=\frac{T_{0}}{2}+(1-T_{0})\omega~.
\end{equation}

To derive the QFI, we first compute the fidelity between $\rho_{1}%
:=\rho_{\alpha}(\tau,T_{0},\omega)$ and $\rho_{2}:=\rho_{\alpha}(\tau
+d\tau,T_{0},\omega)$. These two states have the same CM $V_{1}=V_{2}%
=a^{\prime}\mathbf{I}$, and their mean values differ by $\delta=\sqrt{T_{0}%
}(\sqrt{\tau+d\tau}-\sqrt{\tau})\mathbf{\bar{x}}$. Using the formula of
Ref.~\cite{Benki}, it is straightforward to compute their fidelity
\begin{align}
F(\rho_{1},\rho_{2})  &  =\exp\left[  -\frac{1}{4}\delta^{T}(V_{1}+V_{2}%
)^{-1}\delta\right] \\
&  =\exp\left[  -\frac{T_{0}}{4a^{\prime}}(\sqrt{\tau+d\tau}-\sqrt{\tau}%
)^{2}\bar{n}\right]  .
\end{align}
Using the latter expression in
\begin{equation}
H(\tau)=\frac{8\left[  1-F(\rho_{1},\rho_{2})\right]  }{d\tau^{2}}~,
\end{equation}
and expanding in $d\tau$, we derive the following expression of the quantum
Fisher information (QFI)%
\begin{equation}
H(\tau)=\frac{T_{0}}{T_{0}+2(1-T_{0})\omega}\frac{\bar{n}}{\tau}+O(d\tau)~.
\end{equation}

\section{Correlated-thermal sources\label{app3}}

\subsection{Characterization}

First of all, let us construct the correlated-thermal source. We start from
two single-mode thermal states, $\rho_{H}$ and $\rho_{L}$, with mean number of
photons equal to $\bar{n}_{H}$ and $\bar{n}_{L}$, respectively (with $\bar
{n}_{H}>\bar{n}_{L}$). These states have zero mean and CMs
\begin{equation}
\mathbf{V}_{H}=\mu_{H}\mathbf{I},~~~\mathbf{V}_{L}=\mu_{L}\mathbf{I},
\end{equation}
where $\mu_{H(L)}=\bar{n}_{H(L)}+1/2$. These states are taken as input of a
beam-splitter of transmissivity $\eta$. At the output modes, $A$ and $B$, we
have a Gaussian state with zero mean and CM%
\begin{equation}
\mathbf{V}_{AB}=\left(
\begin{array}
[c]{cc}%
a\mathbf{I} & c\mathbf{I}\\
c\mathbf{I} & b\mathbf{I}%
\end{array}
\right)  ,
\end{equation}
where%
\begin{align}
a  &  :=\eta\mu_{H}+(1-\eta)\mu_{L}~,b:=\eta\mu_{L}+(1-\eta)\mu_{H}~,\\
c  &  :=\sqrt{\eta(1-\eta)}(\mu_{L}-\mu_{H})\text{~}.
\end{align}
In our study we fix $\bar{n}=a-1/2=\eta\bar{n}_{H}+(1-\eta)\bar{n}_{L}$ to
some low value, and we change $\eta$ to create the desired asymmetry between
modes $A$ and $B$.

Note that the correlations are of the negative type $c<0$. Their
(unrestricted) quantum discord can be easily quantified, since it coincides
with their Gaussian discord, according to the results of
Ref.~\cite{OptimalityDiscord}.

\subsection{Evolution}

This correlated-thermal source is sent to probe the lossy channel
$\mathcal{E}_{\tau}$ in the presence of general Gaussian decoherence
$\mathcal{D}$. To model the latter, let us assume a more general scenario than
that of Fig.~\ref{basicPIC}, where the two ancillary modes $E_{1}$ and $E_{2}$
may have different thermal noise, $\omega_{1}$ and $\omega_{2}$. In other
words, we consider an environment described by a Gaussian state with general
CM
\begin{equation}
\mathbf{V}_{E_{1}E_{2}}=\left(
\begin{array}
[c]{cc}%
\omega_{1}\mathbf{I} & \mathbf{G}\\
\mathbf{G} & \omega_{2}\mathbf{I}%
\end{array}
\right)  ,~~\mathbf{G}=\left(
\begin{array}
[c]{cc}%
g & \\
& g^{\prime}%
\end{array}
\right)  .
\end{equation}

In order to be a physical state, this CM must satisfy a set of constraints,
given by~\cite{CMbona,CMbona2}
\begin{equation}
\left\vert g\right\vert <\sqrt{\omega_{1}\omega_{2}},~~\left\vert g^{\prime
}\right\vert <\sqrt{\omega_{1}\omega_{2}},~~\nu^{2}\geq\frac{1}{4}%
\end{equation}
where%
\begin{equation}
\nu^{2}:=\frac{\Delta-\sqrt{\Delta^{2}-4\det\mathbf{V}_{E_{1}E_{2}}}}%
{2},~~\Delta=\omega_{1}^{2}+\omega_{2}^{2}+2gg^{\prime}~.
\end{equation}
Then, we have a separable state if we also impose $\tilde{\nu}^{2}\geq1/4$,
where%
\begin{equation}
\tilde{\nu}^{2}:=\frac{\tilde{\Delta}-\sqrt{\tilde{\Delta}^{2}-4\det
\mathbf{V}_{E_{1}E_{2}}}}{2},~~\tilde{\Delta}=\omega_{1}^{2}+\omega_{2}%
^{2}-2gg^{\prime}~.
\end{equation}
We find that, in the specific case where $g=g^{\prime}$, the condition%
\begin{equation}
\left\vert g\right\vert =\frac{\sqrt{(2\omega_{1}-1)(2\omega_{2}-1)}}{2}%
\end{equation}
guarantees that the state is both physical and separable.

After the action of the lossy channel and the Gaussian environment, the output
Gaussian state $\rho_{AB}^{\text{out}}(\tau)=\mathcal{D}\circ(\mathcal{E}%
_{\tau}\otimes\mathcal{I})(\rho_{AB})$ has zero mean and CM%
\begin{equation}
\mathbf{V}_{AB}^{\text{out}}(\tau)=\left(
\begin{array}
[c]{cccc}%
\tilde{a} &  & c_{1} & \\
& \tilde{a} &  & c_{2}\\
c_{1} &  & \tilde{b} & \\
& c_{2} &  & \tilde{b}%
\end{array}
\right)  , \label{CMapp}%
\end{equation}
where
\begin{align}
\tilde{a}  &  :=T_{0}\tau a+\frac{T_{0}(1-\tau)}{2}+(1-T_{0})\omega_{1},\\
\tilde{b}  &  :=T_{0}b+(1-T_{0})\omega_{2},\\
c_{1}  &  :=T_{0}\sqrt{\tau}c+(1-T_{0})g,~c_{2}:=T_{0}\sqrt{\tau}%
c+(1-T_{0})g^{\prime}\text{~}.
\end{align}

\subsection{Numerical computation of the quantum Fisher information}

To derive the QFI, we first compute the quantum fidelity between the two
(zero-mean) Gaussian states $\rho_{AB}^{\text{out}}(\tau)$ and $\rho
_{AB}^{\text{out}}(\tau+d\tau)$. Following the notation of Ref.~\cite{Benki},
we re-arrange the CM~(\ref{CMapp}) according the ordering $\hat{q}_{A},\hat
{q}_{B},\hat{p}_{A}$ and $\hat{p}_{B}$, so that%
\begin{equation}
\mathbf{V}_{AB}^{\text{out}}(\tau)=\left(
\begin{array}
[c]{cc}%
\tilde{a} & c_{1}\\
c_{1} & \tilde{b}%
\end{array}
\right)  \oplus\left(
\begin{array}
[c]{cc}%
\tilde{a} & c_{2}\\
c_{2} & \tilde{b}%
\end{array}
\right)  .
\end{equation}
Setting $\mathbf{V}_{1}=\mathbf{V}_{AB}^{\text{out}}(\tau)$ and $\mathbf{V}%
_{2}=\mathbf{V}_{AB}^{\text{out}}(\tau+d\tau)$, we compute the auxiliary
matrix
\[
\mathbf{V}_{\text{aux}}=\boldsymbol{\Omega}^{T}(\mathbf{V}_{1}+\mathbf{V}%
_{2})^{-1}\left(  \frac{\boldsymbol{\Omega}}{4}+\mathbf{V}_{2}%
\boldsymbol{\Omega}\mathbf{V}_{1}\right)  ,
\]
where $\boldsymbol{\Omega}:=%
\begin{pmatrix}
0 & \mathbf{I}\\
-\mathbf{I} & 0
\end{pmatrix}
$ is the symplectic form. Finally, the quantum fidelity is given
by~\cite{Benki}%
\[
F(\tau)=\sqrt[4]{\frac{\det\left[  2\left(  \sqrt{\mathbf{I}+\frac{1}%
{4}(\mathbf{V}_{\text{aux}}\boldsymbol{\Omega})^{-2}}+\mathbf{I}\right)
\mathbf{V}_{\text{aux}}\right]  }{\det(\mathbf{V}_{1}+\mathbf{V}_{2})}}.
\]
The latter expression can be expanded for small $d\tau$\ and replaced in the
following formula for the QFI%
\begin{equation}
H(\tau)=\frac{8\left[  1-F(\tau)\right]  }{d\tau^{2}}~.
\end{equation}
With this approach we can numerically derive the curves shown in the figures
of the main text.

\section{Asymmetrically-correlated environment\label{app4}}

For the sake of completeness, we consider here an example where the
environment is correlated but its ancillas introduce different values of
thermal noise. The scenario coincides with that of Fig.~\ref{basicPIC} of the
main text, but now we allow for a more general Gaussian state for the
environment, where the modes $E_{1}$ and $E_{2}$ have thermal noise
$\omega_{1}$ and $\omega_{2}$, respectively. We then assume that the
environmental state is separable with correlations of the type
\begin{equation}
g=g^{\prime}=\frac{-\sqrt{(2\omega_{1}-1)(2\omega_{2}-1)}}{2}~.
\label{gchoice}%
\end{equation}
\begin{figure}[h]
\vspace{-0.2cm}
\par
\begin{center}
\includegraphics[width=0.42\textwidth] {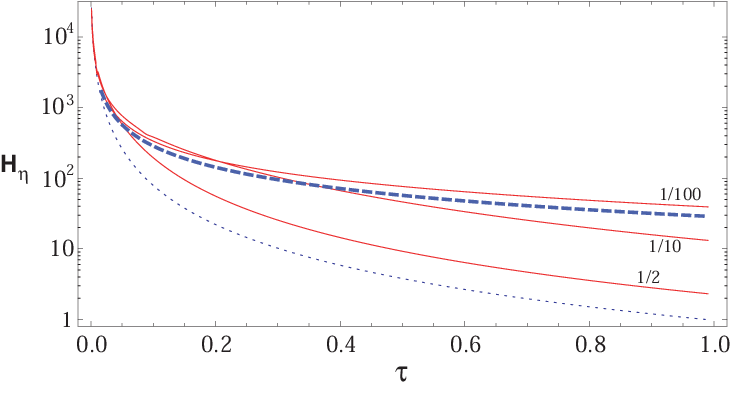}
\end{center}
\par
\vspace{-0.7cm}\caption{Quantum Fisher information $H_{\eta}$ versus
transmissivity $\tau$ for probes irradiating $\bar{n}=50$ signal photons. We
plot the performances of the correlated-thermal source with $\bar{n}_{L}%
\simeq8.3\times10^{-3}$ and having $\eta=1/2$, $1/10$ and $1/100$ (solid red
lines); larger $H_{\eta}$ indicates better precision. These are compared
with\ the single-mode thermal state (dotted blue line) and the coherent-state
benchmark (dashed blue line). Here we consider $T_{0}=0.8$, $\omega_{1}%
=1+1/2$, $\omega_{2}=100+1/2$, and $g=g^{\prime}<0$ as in Eq.~(\ref{gchoice}%
).}%
\label{PIC4_app}%
\end{figure}

\begin{figure}[h]
\vspace{0cm}
\par
\begin{center}
\includegraphics[width=0.42\textwidth] {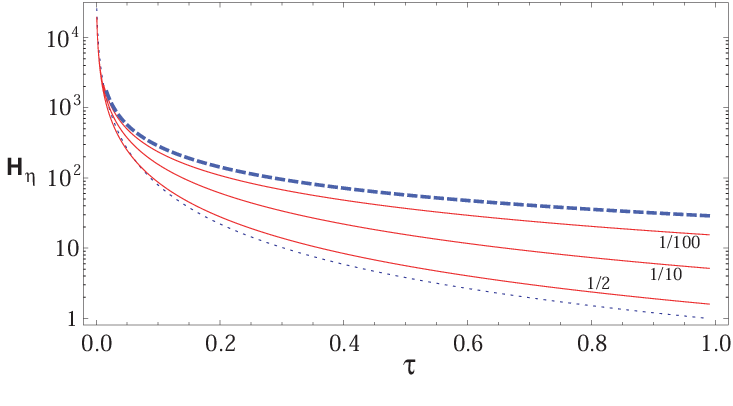}
\end{center}
\par
\vspace{-0.7cm}\caption{Quantum Fisher information $H_{\eta}$ versus
transmissivity $\tau$ for probes irradiating $\bar{n}=50$ signal photons. As
in Fig.~\ref{PIC4_app} but taking positive correlations $g=g^{\prime}%
=\sqrt{(2\omega_{1}-1)(2\omega_{2}-1)}/2$.}%
\label{PIC5_app}%
\end{figure}

From Fig.~\ref{PIC4_app} we see that, for strongly asymmetric noise
$\omega_{2}\gg\omega_{1}$, the coherent-state benchmark can only be beaten by
a sufficiently asymmetric thermal source ($\eta=1/100$), which sends the
majority of the photons through the noisier channel. It is important to remark
that the classical benchmark is outperformed because the \textquotedblleft
negative type\textquotedblright\ of correlations in the environment
($g=g^{\prime}<0$) tend to sustain the \textquotedblleft negative
type\textquotedblright\ of correlations in the input thermal source ($c<0$).
If the environment has the \textquotedblleft positive type\textquotedblright%
\ of correlations ($g=g^{\prime}>0$), then there is a destructive effect and
the classical benchmark is not beaten. See Fig.~\ref{PIC5_app}%
.


\section{Practical receiver designs\label{APP_receivers}}

Here we consider explicit receiver designs based on photon counting, homodyne
and heterodyne detection. We obtain analytical solutions for $\bar{n}_{L}=0$,
$\omega_{1}=\omega_{2}=1/2$, and $g=g^{\prime}=0$, in which case the CM in
Eq.~(\ref{CMapp}) reads
\begin{equation}
\mathbf{V}_{AB}^{\mathrm{out}}(\tau)=\left(
\begin{array}
[c]{cc}%
\frac{1}{2}+\tau\eta n_{H}^{\prime}\mathbf{I} & -n_{H}^{\prime}\sqrt{\tau
\eta(1-\eta)}\mathbf{I}\\
-n_{H}^{\prime}\sqrt{\tau\eta(1-\eta)}\mathbf{I} & \frac{1}{2}+(1-\eta
)n_{H}^{\prime}\mathbf{I}%
\end{array}
\right)  \,, \label{CManal}%
\end{equation}
and we have defined $n_{H}^{\prime}:=T_{0}\bar{n}_{H}$.

\textit{Photon-counting.} The symplectic eigenvalues of the CM in
Eq.\ (\ref{CManal}) are $\frac{1}{2}$ and $\frac{1}{2}+(1+\tau\eta-\eta
)n_{H}^{\prime}$. This means that the state $\rho_{AB}^{\mathrm{out}}(\tau)$
can be written as tensor product, $\rho_{AB}^{\mathrm{out}}(\tau
)=\rho_{A^{\prime}}^{\mathrm{out}}\otimes\rho_{B^{\prime}}^{\mathrm{out}}%
(\tau)$, where $\rho_{A^{\prime}}^{\mathrm{out}}$ is the vacuum state of the
mode $A^{\prime}$ and $\rho_{B^{\prime}}^{\mathrm{out}}(\tau)$ is a thermal
state of the mode $B^{\prime}$ with $m=(1+\tau\eta-\eta)n_{H}^{\prime}$ mean
photons. Denote as $A,A^{\dag}$ and $B,B^{\dag}$ the canonical creation and
annihilation operators of the original pair of modes, and denote as
$A^{\prime},{A^{\prime}}^{\dag}$ and $B^{\prime},{B^{\prime}}^{\dag}$ those of
the new pair of modes. One can easily check that
\begin{equation}
A^{\prime}=(\sin{\theta)}\,A-(\cos{\theta)}\,B,~B^{\prime}=(\cos{\theta
)}\,A+(\sin{\theta)}\,B,
\end{equation}
with
\begin{equation}
\cos{\theta}=-\sqrt{\frac{\tau\eta}{1+\tau\eta-\eta}},~\sin{\theta}%
=\sqrt{\frac{1-\eta}{1+\tau\eta-\eta}}.
\end{equation}
Therefore, the state $\rho_{B^{\prime}}^{\mathrm{out}}(\tau)$ can be written
as
\begin{equation}
\rho_{AB}^{\mathrm{out}}(\tau)=\frac{1}{1+m}\sum_{k=0}^{\infty}\left(
\frac{m}{1+m}\right)  ^{k}|\psi_{k}\rangle\langle\psi_{k}|,\label{rhoB}%
\end{equation}
where $\psi_{k}$ is the state with $k$ photons in the mode $B^{\prime}$,
i.e.,
\begin{align}
|\psi_{k}\rangle &  =\frac{1}{\sqrt{k!}}\left(  \cos{\theta}\,A^{\dag}%
+\sin{\theta}\,B^{\dag}\right)  ^{k}|0\rangle\\
&  =\frac{1}{\sqrt{k!}}\sum_{j=0}^{k}{\binom{k}{j}}\left(  \cos{\theta
}\,A^{\dag}\right)  ^{j}\left(  \sin{\theta}\,B^{\dag}\right)  ^{k-j}%
|0\rangle\\
&  =\frac{1}{\sqrt{k!}}\sum_{j=0}^{k}{\binom{k}{j}}\sqrt{j!(k-j)!}\left(
\cos{\theta}\right)  ^{j}\left(  \sin{\theta}\right)  ^{k-j}|j,k-j\rangle\\
&  =\sum_{j=0}^{k}\sqrt{{\binom{k}{j}}}\left(  \cos{\theta}\right)
^{j}\left(  \sin{\theta}\right)  ^{k-j}|j,k-j\rangle\,.
\end{align}
From the expression in Eq.\ (\ref{rhoB}) we can easily compute the joint
probability of measuring $N_{A}$ and $N_{B}$ photons, which is
\begin{align}
&  P(N_{A},N_{B})\nonumber\\
&  =\frac{1}{1+m}\left(  \frac{m}{1+m}\right)  ^{N_{A}+N_{B}}{\binom
{N_{A}+N_{B}}{N_{A}}}\left(  \cos{\theta}\right)  ^{2N_{A}}\left(  \sin
{\theta}\right)  ^{2N_{B}}\\
&  =\frac{1}{1+m}\left(  \frac{n_{H}^{\prime}}{1+m}\right)  ^{N_{A}+N_{B}%
}{\binom{N_{A}+N_{B}}{N_{A}}}\left(  \tau\eta\right)  ^{N_{A}}\left(
1-\eta\right)  ^{N_{B}}\,.
\end{align}
To compute the (classical) Fisher information associated to this measurement
(photon-counting) we first compute the logarithmic derivative of
$P(N_{A},N_{B})$ with respect to $\tau$, getting
\begin{align}
&  \frac{\partial\log{P(N_{A},N_{B})}}{\partial\tau}\nonumber\\
&  =\frac{1}{\tau}\frac{N_{A}[1+(1-\eta)n_{H}^{\prime}]-(1+N_{B})\tau\eta
n_{H}^{\prime}}{1+n_{H}^{\prime}(1+\tau\eta-\eta)}%
\end{align}
and then we derive the Fisher information
\begin{align}
\mathrm{FI} &  =\langle\left[  \frac{\partial\log{P(N_{A},N_{B})}}%
{\partial\tau}\right]  ^{2}\rangle\\
&  =\frac{1+(1-\eta)n_{H}^{\prime}}{1+(1+\tau\eta-\eta)n_{H}^{\prime}}%
\,\frac{\eta n_{H}^{\prime}}{\tau}\,.
\end{align}
In terms of $\bar{n}=\eta n_{H}=\eta\bar{n}_{H}=\eta n_{H}^{\prime}/T_{0}$ we
obtain the scaling
\begin{equation}
\mathrm{FI}=\frac{\gamma_{\text{phc}}\bar{n}}{\tau},~\gamma_{\text{phc}%
}:=\frac{T_{0}+(1-\eta)T_{0}^{2}\bar{n}\eta^{-1}}{1+(1+\tau\eta-\eta)T_{0}%
\bar{n}\eta^{-1}}.
\end{equation}
Therefore, we recover the same behaviour (up to a small correction) of the
coherent-state benchmark (ideally this benchmark is achieved in the limit
$\eta\rightarrow0$ with $n$ kept constant).


\textit{Homodyne and heterodyne detection.} Consider, for example, that the
$q$ quadratures of the two modes $A$ and $B$ are detected. The outcomes of the
measurements are two correlated Gaussian variables $q_{A}$, $q_{B}$ with joint
probability density $G(q_{A},q_{B})$ with CM
\begin{equation}
V_{q}=\left(
\begin{array}
[c]{cc}%
\frac{1}{2}+\tau\eta n_{H}^{\prime} & -n_{H}^{\prime}\sqrt{\tau\eta(1-\eta)}\\
-n_{H}^{\prime}\sqrt{\tau\eta(1-\eta)} & \frac{1}{2}+(1-\eta)n_{H}^{\prime}%
\end{array}
\right)  .
\end{equation}
The (classical) Fisher information of these correlated Gaussian variables
reads:
\begin{equation}
\mathrm{FI}=\int dq_{A}dq_{B}G(q_{A},q_{B})\left[  \frac{\partial\log
{G(q_{A},q_{B})}}{\partial\tau}\right]  ^{2}.
\end{equation}
Instead of computing this integral directly, we can exploit unitary invariance
and work in the modes $A^{\prime}$, $B^{\prime}$ in which the state becomes a
direct product and the CM is diagonal with eigenvalues $\frac{1}{2}$ and
$\frac{1}{2}+(1+\tau\eta-\eta)n_{H}^{\prime}$. We then obtain
\begin{align}
\mathrm{FI} &  =\int dq_{B}G(q_{B})\left[  \frac{\partial\log{G(q_{B})}%
}{\partial\tau}\right]  ^{2}\\
&  =\frac{1}{2}\left[  \frac{\eta n_{H}^{\prime}}{\frac{1}{2}+(1+\tau\eta
-\eta)n_{H}^{\prime}}\right]  ^{2}%
\end{align}
Similarly, for heterodyne detection we obtain the following expression for the
CM:
\begin{equation}
\mathrm{FI}=\left[  \frac{\eta n_{H}^{\prime}}{1+(1+\tau\eta-\eta
)n_{H}^{\prime}}\right]  ^{2}\,.
\end{equation}
In conclusion, we notice that both homodyne and heterodyne detection are far
from being optimal measurements.

\end{document}